# MODELISATION DES EFFETS DE LA RUGOSITE SUR L'ETUDE DE L'HUMIDITE DES SOLS PAR RADIOMETRIE MICRO-ONDES. APPLICATION A LA MISSION SPATIALE SMOS.

[1]F. Demontoux, [1,2]H. Lawrence et [2]J.P. Wigneron

[1]Université Bordeaux 1 – Laboratoire IMS UMR 5218 – département MCM – 16 avenue Pey-Berland 33607 Pessac - France

francois.demontoux@ims-bordeaux.fr ; heather.lawrence@ims-bordeaux.fr

[2]INRA-unité de Bioclimatologie – EPHYSE, BP 81, Villenave d'Ornon Cedex 33883 - France

jpwigner@bordeaux.inra.fr

**Résumé :** Dans le cadre de la mission SMOS (Soil Moisture and Ocean Salinity) les laboratoires IMS et EPHYSE sont parties prenantes dans l'étude et la validation de l'algorithme d'inversion LMEB. Dans le cadre de cette étude, nous avons développé un modèle radiatif numérique utilisant le logiciel HFSS-ANSOFT. Il permet de calculer l'émissivité de systèmes multicouches (terre-litière par exemple). Nous avons approfondi notre démarche afin d'intégrer de nouveaux paramètres qui peuvent avoir un effet non négligeable sur les mesures. En effet, jusqu'à présent l'effet de la rugosité du sol associé à celui de la litière n'a pas été étudié. De plus, l'épaisseur de litière n'est jamais constante et il faut donc introduire un profil de variation d'épaisseur de litière réaliste. Cet article présente notre travail afin d'intégrer ces profils dans notre modèle numérique. Ces profils peuvent provenir de mesures (profils d'épaisseurs de litière ou de rugosité du sol) ou de calculs (profils de rugosité). Dans les deux cas les données d'entrées de notre modèle sont des fichiers de points XYZ représentant notre profil.

**Mots clés:** Coefficient bistatique, émissivité, modèle radiatif numérique, rugosité, SMOS.

## 1. Introduction

La mission SMOS (Soil Moisture and Ocean Salinity) [1] [2] [3], dont le lancement est prévu pour 2009, consiste au lancement d'un satellite qui transportera un radiomètre interférométrique 2D. Ce dernier effectuera la 1$^e$ cartographie à l'échelle planétaire de l'humidité des sols et de la salinité des océans et ce grâce à un unique instrument de mesure capable de capture d'images des radiations micro ondes émises autour de 1.4GHz.

Le laboratoire IMS (Intégration du Matériau au Système) est parti prenante avec le laboratoire EPHYSE de l'INRA dans l'étude et la validation de l'algorithme d'inversion des données LMEB de la mission SMOS qui liera l'émissivité mesurée à l'humidité des sols. Les mesures de SMOS porteront sur des milieux très différents. Nous nous sommes intéressés aux forêts qui sont présentes dans une majorité des pixels en zone tropicale, boréale et tempérée. Les forêts sont des couverts relativement opaques, sur lesquels le suivi de l'humidité reste problématique. En particulier, l'effet de la litière a, jusqu'ici, été négligé. Ainsi nos études se sont focalisées sur les structures bi-couches sol-litière.

Le but de nos recherches a été dans un premier temps de réussir à mesurer les propriétés diélectriques d'un type de litière et de terre afin d'intégrer ces valeurs à un modèle numérique multi couches de sol que nous avons développé (utilisation du logiciel HFSS de la société ANSOFT [4]). Nous avons ainsi mis en évidence les effets de cette strate sur le système multi couche global. Ceci nous a permis de développer une formulation analytique simple d'un modèle de litière qui pourrait être intégré à l'algorithme de calcul de SMOS.

Il est maintenant nécessaire d'approfondir notre démarche afin d'intégrer de nouveaux paramètres qui peuvent avoir un effet non négligeable sur les mesures. En effet, jusqu'à présent l'effet de la rugosité du sol associé à celui de la litière n'a pas été étudié. De plus, l'épaisseur de litière n'est jamais constante et il faut donc introduire un profil de variation d'épaisseur de litière réaliste. Cet article présente notre travail afin d'intégrer ces profils dans notre modèle numérique. Ces profils peuvent provenir de mesures (profils d'épaisseurs de litière ou de rugosité du sol) ou de calculs (profils de rugosité). Dans les deux cas les données d'entrées de notre modèle sont des fichiers de points XYZ représentant notre profil. L'introduction de ces profils est un problème délicat dans le cas de la méthode de résolution des équations de Maxwell des éléments finis que nous utilisons pour notre modèle numérique. En effet, contrairement à la méthode des différences finies (FDTD), par exemple, nous travaillons sur un maillage de type tétraédrique sur lequel l'introduction d'un profil réel est difficile.

## 2. Méthode d'introduction de la rugosité

### 2.1 Présentation de la méthode

La méthode d'introduction des profils de rugosité dans le logiciel HFSS peut être résumée en quatre phases (figure 1). Lors de la première phase (A et A' sur la figure 1) un fichier de points xyz représentant la surface rugueuse à simuler est créé ou obtenu par mesures in situ. Les paramètres utilisés pour définir la rugosité sont kσ et kl (k=2π/λ). σ est l'écart type des hauteurs et l la longueur d'auto-corrélation. Dans cette étude, la fonction d'auto-corrélation est de type gaussienne.

Puis un fichier géométrie 2D représentant l'interface est créé (phase B sur la figure 1 et représentation figure 2). Ce fichier géométrie 2D (*.step) est ensuite importé dans un logiciel de dessin 3D (ModelShop model design). Un objet géométrique 3D (*.sat) est alors créé (phase C figure 1). Enfin, ce fichier est importé sous HFSS et reconnu comme un objet 3D à qui nous pouvons par exemple affecter des propriétés diélectriques (figure 3).

Il est alors possible pour le « mailleur » automatique du logiciel de découper le volume créé en tétraèdres (figure 4). Ce découpage servira à l'algorithme de calcul du champ électromagnétique et permettra l'évaluation du champ diffracté par la structure.

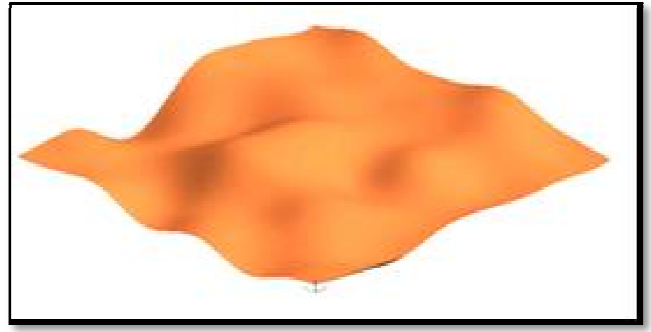

**Figure 2.** *Définition du fichier de géométrie 2D représentant la rugosité d'une interface*

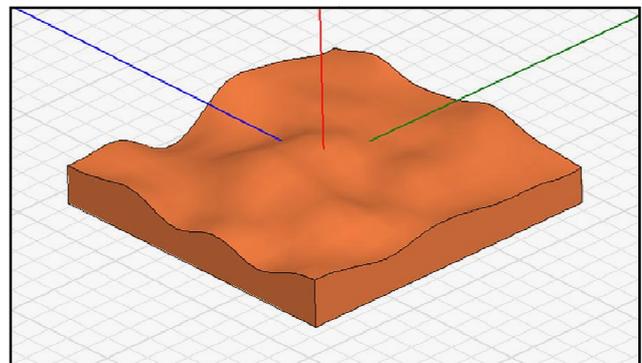

**Figure 3.** *Définition du fichier géométrie 3D représentant la strate.*

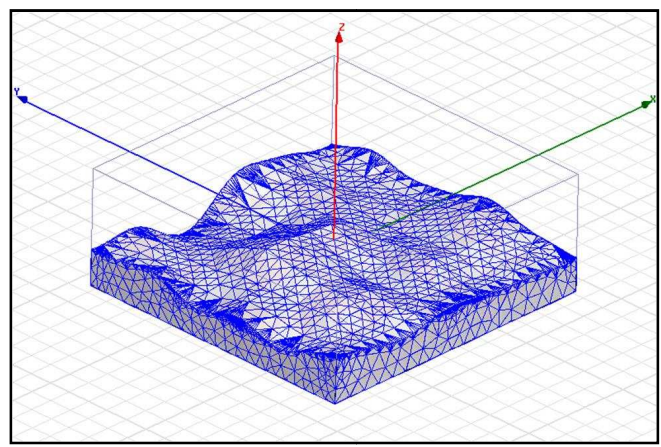

**Figure 4.** Maillage par le logiciel HFSS de la géométrie 3D représentant une strate.

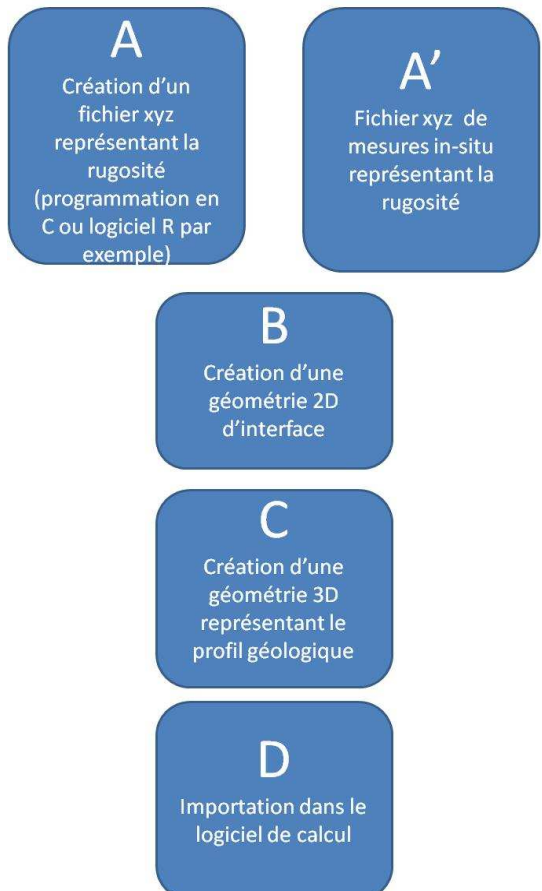

**Figure 1.** Principe de la méthode de prise en compte de la rugosité.

## 3. Résultats

### 2.1 Calcul du champ électrique.

Le logiciel HFSS permet le calcul en champ proche et en champ lointain du champ diffracté par notre structure géologique. Des outils de post-traitement des résultats permettent de visualiser de nombreux paramètres (exemple figure 5)

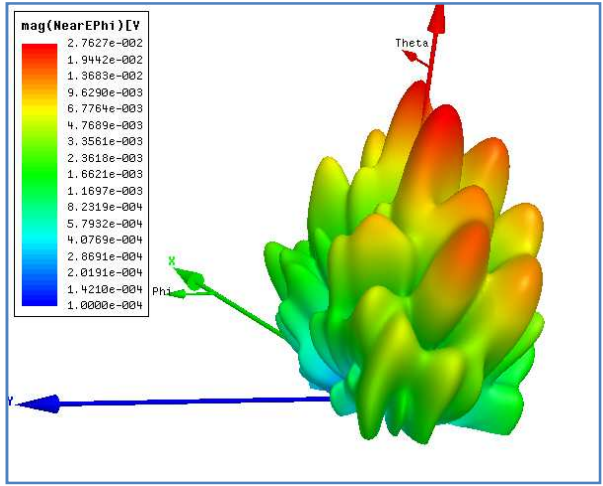

**Figure 5.** *Amplitude du champ électrique diffracté par une surface rugueuse- kl=6.28, k$\sigma$=1, $\varepsilon$=7-j0.8*

### 2.1 Calcul du coefficient bi-statique

En télédétection active, le coefficient bi-statique est une grandeur très importante pour caractériser les structures géologiques et pour caractériser leur rugosité en particulier. Il est défini à l'aide de l'équation suivante [5] :

$$\sigma_{rt}^0(\theta_s, \phi_s; \theta, \phi) = \frac{4\pi R^2 |E_r^s|^2}{A|E_t^i|^2} \quad (1)$$

Avec $\sigma_{rt}^0(\theta_s, \phi_s; \theta, \phi)$ le coefficient bi-statique ;
$E_r^s$ le champ diffracté ;
$E_t^i$ le champ incident ;
R : distance sol-observateur
A : surface éclairée

Ce coefficient est fonction de l'angle d'incidence et de l'angle d'observation choisie mais aussi de la polarisation, de la nature des sols (permittivités, multi couches …) ou de la rugosité des interfaces. Une observation radar d'une telle structure va nous permettre d'obtenir un coefficient global représentant une moyenne de ce coefficient sur l'ensemble de la surface éclairée. En simulation nous devons limiter la taille de la structure géologique. Pour obtenir un coefficient bi-statique représentatif de ce que nous observerions dans la réalité nous devons créer et simuler de nombreuses surfaces de même rugosité et de profils différents. Une moyenne des courbes obtenues nous permet d'avoir ce coefficient bi-statique « global » (figure 5).

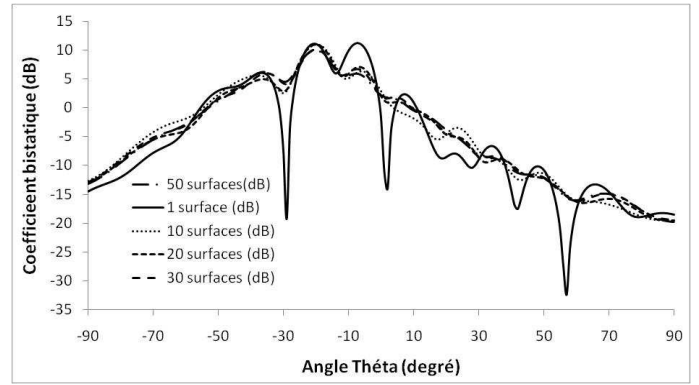

**Figure 6.** Co*efficient bi-statique pour différentes surfaces de même rugosité*
*kl=6.28, k$\sigma$=1, $\varepsilon$=7-j0.8*
*Théta incident = 20 °, Phi = 0°*
*Polarisation HH*

Sur la figure 6 nous pouvons noter que la courbe résultante obtenue ne varie plus beaucoup au-delà de 20 surfaces utilisées.

Ainsi nous obtenons en fonction de la polarisation la courbe de variation du coefficient bi-statique (figure 7).

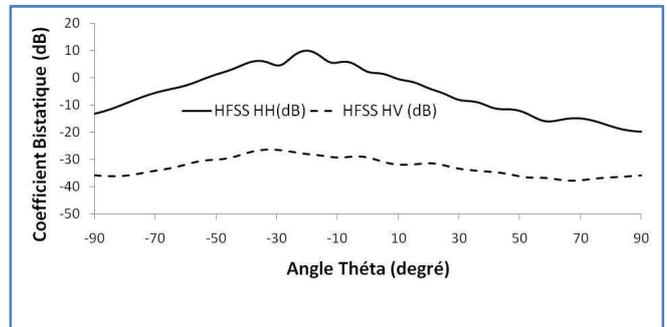

**Figure 7.** Co*efficient bi-statique en prenant 30 surfaces rugueuses*
*kl=6.28, k$\sigma$=1, $\varepsilon$=7-j0.8*
*Théta incident = 20 °, Phi = 0°*
*Polarisation HH et HV*

Ces premiers résultats sont très encourageants et sont en accords avec les résultats issus d'autres méthodes (FDTD, méthode des moments). Un affinage de la précision pourra être obtenu par optimisation du modèle afin d'utiliser moins de ressources informatiques et ainsi pouvoir diminuer les critères de convergence de l'algorithme de calcul.

### 2.3 Etude de l'émissivité

Le calcul du coefficient bi-statique peut conduire au calcul de l'émissivité de la structure [5] (eq. 2, 3 et 4).

$$e_r(\theta, \phi) = 1 - \Gamma_r(\theta, \phi) \quad (2)$$

$$\Gamma_r(\theta, \phi) = \frac{1}{4\pi} \iint [\gamma_{rr}(\theta_s, \phi_s; \theta, \phi) + \gamma_{tr}(\theta_s, \phi_s; \theta, \phi)] d\Omega_s \quad (3)$$

$$\sigma_{rt}^0 = \gamma_{rt} \cos\theta \quad (4)$$

Dans ce cadre nous avons développé un modèle numérique de calcul de l'émissivité de structures géologiques complexes [6] [7]. Ce modèle doit permettre de créer une base de données simulées qui permettront de valider l'algorithme d'inversion de la mission spatiale SMOS. Ce modèle permet de calculer l'émissivité des structures par intégration de leur coefficient bi-statique (eq 2, 3, 4 [5]). Un exemple de résultats est présenté (figure 8). Notre méthode de calcul de l'émissivité est en cours de validation par comparaison avec d'autres méthodes [8] (AIEM, MoM).

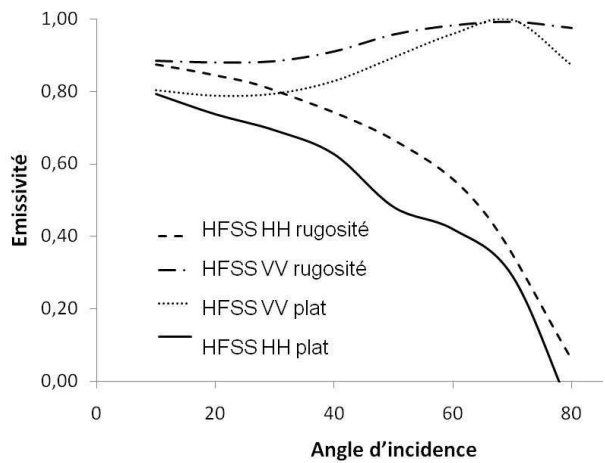

**Figure 8 :** E*missivité d'une couche de terre plane ou rugueuse*
*rugosité : kl=6.28, kσ=1*
*ε=7-j0.8*

Ces premiers résultats mettent l'accent sur l'importance de la prise en compte des rugosités d'interfaces dans le calcul de l'émissivité de structures géologiques. En plus de l'effet observé sur le niveau d'émissivité nous pouvons noter que la rugosité réduit l'écart entre les valeurs issues des deux polarisations HH et VV.

## 5. Conclusion et perspectives

Nous avons mis au point et utilisé notre méthode d'introduction de rugosité dans le logiciel HFSS. Les premiers résultats obtenus sur le coefficient bi-statique et l'émissivité sont très encourageants. Cette démarche permettra l'étude de la réponse électromagnétique de structures géologiques complexes dans le cadre de projets spatiaux de télédétection active ou passive. A titre d'exemple ,elle va permettre d'introduire des configurations de sols Martiens réalistes dans un modèle de simulation électromagnétique du fonctionnement du radar WISDOM embraqué sur le rover de la mission Exomars du projet Aurora de l'ESA [9][10].

## Références